 \documentclass[authoryear,preprint,review,12pt]{elsarticle}
\usepackage[colorlinks=true,linkcolor=black, citecolor=blue, urlcolor=blue]{hyperref}

		\usepackage{amssymb}
		\newif\ifincludegraphics
		\includegraphicstrue 
		\usepackage{epstopdf}
		\usepackage{multirow}
		\usepackage{tabularx}
		\usepackage{amsmath}
      		\usepackage{footnote}
		\usepackage[table,svgnames,x11names]{xcolor}  
 		\usepackage{siunitx}
\usepackage{lineno}

\journal{Icarus}
\newcommand\phaseNCH{$\bf{\overline{N_{2}}}$:CH$_{4}$}
\newcommand\phaseCHN{$\bf{\overline{CH_{4}}}$:N$_{2}$}
\newcommand\micron{$\mathrm{\mu}$m}
\newcommand\eg{\textit{e.g.}}
\newcommand\methane{CH$_{4}$}
\newcommand\nitrogen{N$_{2}$}
\newcommand\NH{New Horizons}
\newcommand\water{H$_{2}$O}

\begin{document}

\begin{frontmatter}



\title{Pluto's global surface composition through 
pixel-by-pixel Hapke modeling of New Horizons Ralph/LEISA data}


\author[label1]{S. Protopapa}
\author[label2]{W. M. Grundy}
\author[label3]{D.C. Reuter}
\author[label1]{D.P. Hamilton}
\author[label6,label5]{C. M. Dalle Ore}
\author[label4]{J.C. Cook}
\author[label5]{D.P. Cruikshank}
\author[label8]{B. Schmitt}
\author[label8]{S. Philippe}
\author[label8]{E. Quirico}		
\author[label7]{R. P. Binzel}
\author[label7]{A.M. Earle}
\author[label4]{K. Ennico}
\author[label4]{C.J.A. Howett}
\author[label3]{A.W. Lunsford}
\author[label4]{C. B. Olkin}
\author[label4]{A. Parker}
\author[label4]{K.N. Singer}
\author[label4]{A. Stern}	
\author[label10]{A. J. Verbiscer}	
\author[label9]{H. A. Weaver}	
\author[label4]{L.A. Young}	
\author{the New Horizons Science Team}
\address[label1]{University of Maryland, Department of Astronomy, College Park, Maryland 20742, USA}
\address[label2]{Lowell Observatory, Flagstaff, Arizona 86001, USA}
\address[label3]{National Aeronautics and Space Administration Goddard Space Flight Center, Greenbelt, Maryland 20771, USA}
\address[label6]{SETI Institute}
\address[label5]{NASA Ames Research Center}
\address[label4]{Southwest Research Institute, Boulder, Colorado 80302, USA}
\address[label8]{Institut de Plan\'etologie et Astrophysique de Grenoble, UGA / CNRS, IPAG, Grenoble Cedex 9, France}
\address[label7]{Massachusetts Institute of Technology}
\address[label9]{Johns Hopkins University Applied Physics Laboratory}
\address[label10]{University of Virginia}
\begin{abstract}
On July 14th 2015, NASA's New Horizons mission gave us an unprecedented  detailed view of the Pluto system.
The complex compositional diversity of Pluto's encounter hemisphere was revealed by the Ralph/LEISA infrared spectrometer on board of New Horizons. We present compositional maps of Pluto defining the spatial distribution of the abundance and textural properties of the volatiles methane and nitrogen ices and non-volatiles water ice and tholin. These results are obtained by applying a pixel-by-pixel Hapke radiative transfer model to the LEISA scans.  Our analysis focuses mainly on the large scale latitudinal variations of methane and nitrogen ices and aims at setting observational constraints to volatile transport models. Specifically, we find three latitudinal bands: the first, enriched in methane, extends from the pole to 55$^{\circ}$N, the second dominated by nitrogen, continues south to 35$^{\circ}$N, and the third, composed again mainly of methane, reaches 20$^{\circ}$N. We demonstrate that the distribution of volatiles across these surface units can be explained by differences in insolation over the past few decades. The latitudinal pattern is broken by Sputnik Planitia, a large reservoir of volatiles, with nitrogen playing the most important role. The physical properties of methane and nitrogen in this region are suggestive of the presence of a cold trap or possible volatile stratification. Furthermore our modeling results point to a possible sublimation transport of nitrogen from the northwest edge of Sputnik Planitia toward the south.
\end{abstract}


\end{frontmatter}


\section{Introduction}\label{Introduction}
NASA's \NH~mission completed a close approach to the Pluto system on 14
July 2015 reaching a distance of 12,000 km from
the dwarf planet's surface \citep{Stern2015Sci}. A wealth of ground-based data of the Pluto system had been collected prior to the \NH~mission, motivated first by its status as the outermost planet and later by its being among the few trans-Neptunian objects (TNOs) bright enough for detailed studies to further advance our knowledge of the Kuiper Belt. 

The synergy of ground-based observations, modeling efforts and laboratory studies have highlighted over the course of the past years, among other things, 1) the presence of methane (CH$_{4}$), nitrogen (N$_{2}$), carbon monoxide (CO) and ethane (C$_{2}$H$_{6}$) ices on Pluto together with tholins, 2) constraints on Pluto's surface temperature, and 3) the state of CH$_{4}$ ice. CH$_{4}$, \nitrogen, and CO ices were detected through ground-based measurements in the near-infrared wavelength range \citep{Cruikshank1976,Owen1993}. While CH$_{4}$ is the most spectroscopically active constituent among Pluto's ices, with several strong absorption bands, CO and N$_{2}$ were identified by the detection of the bands at 2.35 $\mu$m, as well as the weaker band at 1.58 $\mu$m, and 2.15 $\mu$m, respectively. Models of Pluto's near-infrared reflectance spectra including C$_{2}$H$_{6}$ yield an improved reduced $\chi^{2}$, leading to the conclusion that this ice is also present on the surface of the planet \citep{DeMeo2010,Holler2014,Merlin2015}. The red slope of Pluto's continuum \citep{Bell1979,GrundyFink1996,Lorenzi2016} observed from the visible to around $\sim$1~\micron~has been attributed to the presence of organic materials on the surface of the body, such as tholins. These are the refractory residues obtained from the irradiation of gases and ices containing hydrocarbons \citep{Cruikshank2005}. In the case of Pluto, tholins may form in situ through energetic processing of CH$_{4}$ and \nitrogen~ices \citep{Cruikshank2016}. The spectral profile of the 2.15-$\mu$m N$_{2}$ absorption band is temperature dependent \citep{Grundy1993,Tryka1993} and transitions from broad to very narrow according to whether N$_{2}$ is in the hexagonal $\beta$- (above 35.6 K) or cubic $\alpha$-phase, respectively. \citet{Tryka1994}, using N$_{2}$ as a spectral `thermometer', inferred a surface temperature of 40$\pm$2 K\null. At such temperature, thermodynamic equilibrium dictates that pure CH$_{4}$ ice cannot co-exist with pure N$_{2}$ ice. If both species are present, then instead of pure ices the phases are \citep{Trafton2015}: CH$_{4}$ saturated with N$_{2}$ (\phaseCHN) and N$_{2}$ saturated with CH$_{4}$ (\phaseNCH). The solubility limits of \methane~and \nitrogen~in each other are temperature dependent and are equal to about 5\% (\phaseNCH, with 5\% \methane) and 3\% (\phaseCHN~with 3\% \nitrogen) at 40~K \citep{Prokhvatilov1983}. \methane~when dissolved in \nitrogen~presents
absorption bands shifted toward shorter wavelengths compared
to the central wavelengths of pure \methane~\citep{QuiricoSchmitt1997,Grundy2002CH4}. This shift, which is wavelength dependent and is observed in Pluto spectra \citep{Owen1993}, varies with the \methane~abundance in the mixture: the larger the \methane~concentration
the smaller the blueshift \citep{SchmittQuirico1992,QuiricoSchmitt1997,Protopapa2015}. 

While ground-based measurements have played a remarkable role in the growth of knowledge about Pluto's composition \citep{Cruikshank2015}, they 
were limited in providing composition maps of Pluto. From the ground, it is possible to investigate changes of
Pluto surface composition with longitude by comparing measurements obtained at different rotational phases. Variations with latitude can be determined only by monitoring Pluto as it moves around the Sun during the 248 year orbit. However, over time scales of years, Pluto might undergo a resurfacing process \citep{Stern1988,Buie2010,Buie2010b}. \nitrogen, CO, and \methane~ices are all volatiles at Pluto surface temperature, of which \methane~is the least volatile, and support Pluto's atmosphere. Changes in insolation over the course of a Pluto's orbit can result in the bulk migration of volatile ices \citep{Spencer1997,hansen96,young12,young13,hansen15}. It is therefore difficult to disentangle temporal and spatial variations using ground-based observations \citep{Grundy2013,Grundy2014}. 

\NH~provided a detailed snapshot of the Pluto system with the goal, among others, to map Pluto surface composition and search for additional surface species \citep{Young2008}. \NH~confirmed the presence on Pluto surface of the volatile ices \nitrogen, \methane, and CO and detected the existence of the non-volatile water (\water) and possibly methanol (CH$_{3}$OH) ices \citep{Grundy2016,Cook2016}. The detection of  \water-ice did not come as a surprise given that most TNOs are characterized by \water-ice dominated spectra \citep{Barucci2008}. Furthermore, Triton, which is thought to be a former TNO captured into a retrograde orbit around Neptune \citep{agnor06}, presents a near-infrared spectrum similar to that of Pluto with clear evidence of \water-ice absorption bands \citep{Quirico1999,Cruikshank2000}. Contrary to Triton \citep{Cruikshank1993}, Pluto does not display the signatures of carbon dioxide (CO$_{2}$) ice on the surface \citep{Grundy2016}.  

\NH~revealed striking variations in the distribution of Pluto's ices \citep{Grundy2016}. These results rely on the analysis of spectral parameters, including band depth and equivalent width. However, the main limitation of that approach is the incapability of disentangling relative abundance from grain size effects. This is possible only by means of radiative transfer modeling of the absorption bands over a wide wavelength range, a method that has its own limitations (see Section \ref{Model Discussion}). In this paper, we will discuss constraints on the abundances and scattering properties of the materials across the surface of Pluto, focusing mainly on the distribution of \nitrogen~and \methane~ices and the relation between their distribution and surface geology, which is vital to set observational constraints for volatile transport models \citep{young13,hansen15,Bertrand2016}.

\section{Observations}
Spatially resolved near-infrared spectra of Pluto's surface were acquired using the Linear Etalon Imaging Spectral Array (LEISA), part of the \NH~Ralph instrument \citep{Reuter2008}. LEISA consists of a wedged filter placed in close proximity to a  256$\times$256 pixels detector array. The filter consists of two segments covering the wavelength range 1.25-2.5~\micron~and 2.1-2.25~\micron~at the resolving power ($\lambda$/$\Delta\lambda$) of 240 and 560, respectively. The low-resolution 1.25-2.5~\micron~segment is used to infer the surface composition of Pluto as outer solar system ices such as \nitrogen, \methane, \water~have strong unique absorption bands in this wavelength region. The high-resolution 2.1-2.25~\micron~segment is instead sensitive to the spectral shape of the 2.15-\micron~absorption band of \nitrogen~ice, which is temperature dependent, as well as to the spectral shape and position of the 2.2-\micron~absorption band of \methane~ice, which is important to assess the \methane~to \nitrogen~mixing ratio. The wavelength varies along the row direction of the detector array. LEISA is operated in a scanning mode. A series of image frames, $N$, are acquired while scanning the field of view across the target surface in a push broom fashion. The target moves through the image plane, along the spectral direction, also called along-track direction. A process of co-registration of each wavelength image is applied, removing motion and optical distortions, to obtain a three-dimensional array spectral cube where each frame images the target surface at a distinct LEISA wavelength. The same process of co-registration is applied to generate a wavelength spectral cube, such that each pixel has its own spectral array. This way we take into account an existing spectral distortion (`smile') of 2-3 spectral elements along the 256 scanning cross-track pixels due to a slight curvature induced by the filter deposition process. 

In this paper we present two LEISA resolved scans of Pluto collected at a distance from Pluto's center of $\sim$100,000~km at a spatial scale of 6 and 7~km/pixel (see Table~\ref{Table1} for details). The LEISA data presented here have been calibrated through the most up-to-date pipeline processing, which includes bad pixel masking, flat fielding, and conversion from DN to radiance $I$ expressed in erg s$^{-1}$cm$^{-2}$\AA$^{-1}$str$^{-1}$. The latter is normalized by the incoming solar flux to obtain reflectance ($I/F$). Efforts to improve the radiometric calibration and flat-field are currently underway. Improvements will probably be made as more spectral data from Pluto and Charon are downlinked and become available. The possible impact of calibration changes have been considered in the results presented here. 

The last step in the data processing consists of determining the Pluto latitude and longitude corresponding to each LEISA pixel. Integrated Software for Imagers and Spectrometers (ISIS, \url{https://isis.astrogeology.usgs.gov/}) provided by the United States Geological Survey (USGS) uses
the spacecraft attitude history, which is measured and recorded as part of the standard housekeeping data, along with its LEISA camera model and
the reconstructed spacecraft trajectory to determine where each LEISA
pixel falls on the target body.  We used the ISIS software to perform an
orthographic projection of the LEISA data to a sphere at the target's size
and location relative to the spacecraft as of the mid-time of each scan. The two LEISA scans presented in this paper combined cover Pluto's full disk and they were both projected to a common orthographic
viewing geometry appropriate for the mid-time between the two.
The re-projection was done to a target grid with a spatial scale of 2
km/pixel, a higher resolution than the native LEISA pixel scales (see Table~\ref{Table1}).  The point
of sub-sampling is to minimize degradation of spatial information as
a result of the nearest neighbor re-sampling.  Using the ISIS ``translate'' routine, we applied a global shift of the LEISA data (consistent across all LEISA wavelengths) to correct for a few pixels mismatch between the LEISA cube and the much higher resolution base map obtained with the Long Range Reconnaissance Imager \citep[LORRI,][]{Cheng2008} projected to the
same geometry. We took into account higher order corrections by means of the ISIS ``warp'' routine. This is based on a user supplied control network constructed using features that were recognizable in both LEISA and LORRI data. These corrections resulted in LEISA cubes estimated to be geometrically
accurate to a little better that a single LEISA pixel.  The mid-scan
geometry was taken as a reasonable approximation of the illumination and
viewing geometry for each pixel.  

The modeling analysis presented in this paper is conducted on the LEISA scans degraded to a spatial resolution of $\sim$12 km/pixel through bin averaging. The mean absolute deviation of the incident and emergent angles within a bin is generally less than $\sim$0.2 deg but at the limb where it approaches $\sim$1 deg. The two scans overlap in a region crossing Elliot Crater\footnote{All place names used in this paper are informal designations} and Sputnik Planitia. To assess the error in the I/F data introduced by variations in the quality of the flat-field, scattered light, and other spatially correlated noise sources, we compared the two scans in the region of overlap.  A variety of statistical comparisons all consistently produce a conservative estimate of 5\% as the mean 1$\sigma$ uncertainty contributed by these sources of error.

\begin{table}
\caption{Details of the LEISA scans presented in this paper.}
\resizebox {0.5\textwidth }{!}{%
\begin{tabular}{l c c}
\hline
Request ID &P\_LEISA\_Alice\_2a&P\_LEISA\_Alice\_2b\\
MET&299171897&299172767\\
S/C Start Time, UTC& 2015-07-14&2015-07-14\\
&09:26:19&09:40:49\\
Phase [deg] & 21.7 & 22.4\\
SubSol Lat [deg]&51.6&51.6\\
SubSol Lon [deg]&133.4&132.9\\
SubS/C Lat [deg]&38.8&38.2\\
SubS/C Lon [deg]&158.6&158.7\\
Spatial Scale [km/px]&7&6\\
\hline
\end{tabular}}
\label{Table1}
\end{table}
\section{Spectral Modeling}\label{Spectral Modeling}
The goal of our analysis is to understand the spatial distribution of the abundance and textural properties of each Pluto's surface component. To this end we performed a pixel-by-pixel modeling analysis of the LEISA spectral cubes.  This approach requires the application of the same modeling strategy across the entire visible face of Pluto, so that a systematic and comparative study  between the composition of the different surface units can be conducted. This is at the expense of possible compositional peculiarities in small surface areas. 

We now describe the modeling of a single pixel LEISA spectrum. We use the scattering radiative transfer model of \citet{Hapke1993} to compute the bidirectional reflectance $r$ of a particulate surface as
\begin{multline}\label{eq: bidirectional reflectance NOT isotropic scattering SHOE roughness}
r(i,e,g)=\frac{w}{4\pi}\frac{\mu_{0e}}{\mu_{e}+\mu_{0e}}\{[1+B(g)]P(g)+ \\ H(\mu_{e})H(\mu_{0e})-1\}S(i,e,g).
\end{multline}
The single scattering albedo $w$ is the ratio of the scattering efficiency to the
extinction efficiency ($w = 0$ implies that the particles absorb all the radiation) and it 
can be computed only if the optical
constants $n$ and $k$, the real and imaginary part of the refractive index, respectively, of
the surface component, as a function of the wavelength $\lambda$, are known. We adopt the equivalent
slab model presented by \citet{Hapke1993} to compute $w$, given the complex refractive index. Notice that $w$ is a function of the particle's effective diameter $D$, which is a free parameter in our analysis. The terms $\mu_{0e}$ and $\mu_{e}$ are related to the cosine of the incidence
angle ($i$) and emission angle ($e$), respectively, with additional terms to account for the tilt of the surface, due to surface
roughness. The latter is also accounted for by the shadowing function $S(i,e,g)$, which involves the mean slope angle
 $\overline{\theta}$. The backscatter function $B(g)$ is an approximate expression for the opposition effect and it is given by 
\begin{equation}
B(g) = \frac{B_{0}}{1+(1/h)\tan(g/2)},
\end{equation}
where $g$ is the phase angle, $h$ is the compaction parameter and characterizes the width of the nonlinear
increase in the reflectance phase curve with decreasing phase angle (the opposition surge), and $B_{0}$ is an empirical factor that represents the amplitude of the opposition effect. We adopt for $P(g)$ a single lobe Henyey and Greenstein function
\begin{equation}
P(g) = \frac{1-\xi\:^2}{(1+2\:\xi \cos\:g+\xi^{2})^{3/2}},
\end{equation}
where $\xi$ is the cosine asymmetry factor. The Ambartsumian--Chandrasekhar $H$-functions are computed using the approximation proposed by \citet{Hapke2002}.

Equation (\ref{eq: bidirectional reflectance NOT isotropic scattering SHOE roughness}) can be used to compute the bidirectional reflectance
of a medium composed of closely packed particles of a single component. However, the surface of interest is a mixture of different constituents \citep{Grundy2016}. Even at the pixel level, more than one component can be present. Therefore, in order
to calculate synthetic reflectance spectra for comparison with the observational data, it is necessary to compute the reflectance of a mixture of
different types of particles. We have considered an areal (also called geographical) mixture, which consists of materials of different
composition and/or microphysical properties that are spatially isolated from
one another. We adopt this approach as it is the most simple and it provides satisfactory results. In the case of areal mixture, the bidirectional reflectance spectra of the individual components ($r_{i}$)
are summed with weights equal to the fractional area of
each terrain ($F_{i}$), as shown below
\begin{eqnarray}\label{riflettanza di una mistura areale}
r=\sum_{i}F_{i}r_{i} & \textrm{where}\ \sum_{i}F_{i}=1.
\end{eqnarray}
By considering an areal mixture, we are implying that there is no multiple scattering between different components. Given the definition of bidirectional reflectance by \citet{Hapke1993} as the ratio between the radiance at the detector and the irradiance incident on the medium, we have that $I/F = \pi r$. We set the cosine asymmetry
parameter $\xi\:=\:-0.3$, the compaction parameter $h\:=\:0.5$, the amplitude of the opposition effect $B_{0}\:=\:1$, and  mean roughness slope $\overline\theta\:=\:10^{\circ}$, following previous studies \citep{Olkin2007,Buie2010b}.

The free parameters in our model are effective diameter ($D_{i}$) and contribution of each surface
terrain to the mixture ($F_{i}$) . They are iteratively modified by means of a Levenberg-Marquardt $\chi^{2}$ minimization algorithm until a best-fit to the observations is achieved. 

It is possible to determine a spatial map for the abundance and grain size of each surface material by applying the modeling analysis described above pixel-by-pixel. Notice that we consider the other Hapke parameters to be constant across Pluto's surface.  
\section{Mixture endmembers}
The surface components considered in the mixture are \water~ice, \phaseNCH~ice, \phaseCHN~ice, and Titan tholin. The choice of the surface terrain units relies on the spectral evidence collected over the course of several years, which are outlined in Section \ref{Introduction}. For details about the optical constants used for each surface material see Table~\ref{Table2} and the following discussion.
\begin{table*}
\caption{Optical constants}
\resizebox {\textwidth }{!}{%
\begin{tabular}{l l c c c p{4cm}}
\hline
Material &Source&Temperature&Filename&Wavelength range& Notes\\
                &             & [K]                 &                  & [\micron]&\\
\hline
\\
Titan Tholin & \citet{Khare1984}&&&0.02--920.0&\\
\\
\methane-ice &  \citet{Grundy2002CH4}\footnotemark[1]		&39		&optcte-Vis+NIR+MIR-CH4cr-I-39K	&0.7--5.0&\\
\\
\nitrogen-ice  & \citet{Grundy1993}\footnotemark[1]		&36.5	&optcte-NIR-beta-N2-36.5K		& 2.062--4.762	&\nitrogen~is in $\beta$-phase. We set $n\:=\:1.23$ and $k\:=\:0$, below 2.1~\micron.\\
\\
\nitrogen:\methane-ice& \citet{QuiricoSchmitt1997}\footnotemark[1]&36.5&optcte-NIR-CH4-lowC-beta-N2-36.5K&1--5&$\beta$-\nitrogen:\methane~solid solution with \methane~concentrations lower than 2\%. The absorption coefficient of diluted \methane~is normalized to a concentration of 1.\\
\\
\water-ice & \citet{Grundy1998}\footnotemark[1] &40 & & 0.96-2.74&\\
\hline
\end{tabular}}
\footnotemark[1]{Optical constants are available at \url{http://ghosst.osug.fr/}.}
\label{Table2}
\end{table*}

\subsection{Tholins}
While there is no doubt about the presence of tholins on the surface of Pluto, the specific type of organic material responsible for Pluto's color, ranging from yellow to red \citep{Stern2015Sci,Grundy2016}, is still unknown. Different types of tholins have been studied but there are only a few for which optical constants are available \citep[for a review on the different kind of tholins the reader is refereed to][]{deBergh2008}. In particular,
the most common are Triton tholin and Titan tholin, which are obtained by irradiating
gaseous mixtures of \nitrogen~and \methane. The difference between the two is in the initial \nitrogen~to \methane~gaseous mixing ratio \citep{McDonald1994, Cruikshank2005,deBergh2008}. Spectrally, both types of tholins present a red slope in the visible, but Triton tholin contrary to Titan tholin also displays a red slope in the near-IR. Laboratory measurements are currently underway to obtain refractory tholins particularly
relevant to Pluto (Pluto's ice tholin) through UV and low-energy electron bombardment of a mixture of Pluto's ices \citep[\nitrogen:\methane:CO = 100:1:1,][]{Materese2015ApJ}.  A preliminary reflectance spectrum of Pluto's ice tholin has been presented by \citet{Cruikshank2015DPS,Cruikshank2016}, but no optical constants are available at this time. The spectrum shows a red slope between 0.5 and 1~\micron, and turns blue in the range between 1 and 2.5~\micron. We adopt Titan tholin in the modeling (see Table~\ref{Table2}), given the similar spectroscopic behavior with respect to Pluto's ice tholin. The choice of Titan instead of Triton tholin is further supported by the blue slope of the Pluto/LEISA spectra extracted in the area informally known as Cthulhu Regio, which is a large, dark region along Pluto's equator and one of the darkest and reddest regions on the surface, at visible wavelengths. However, we emphasize that  the search for the most
appropriate tholin continues and is not the goal of this preliminary
study. 
\begin{figure}[!t]
	\ifincludegraphics
 	\centering
	\includegraphics[width=0.45\textwidth]{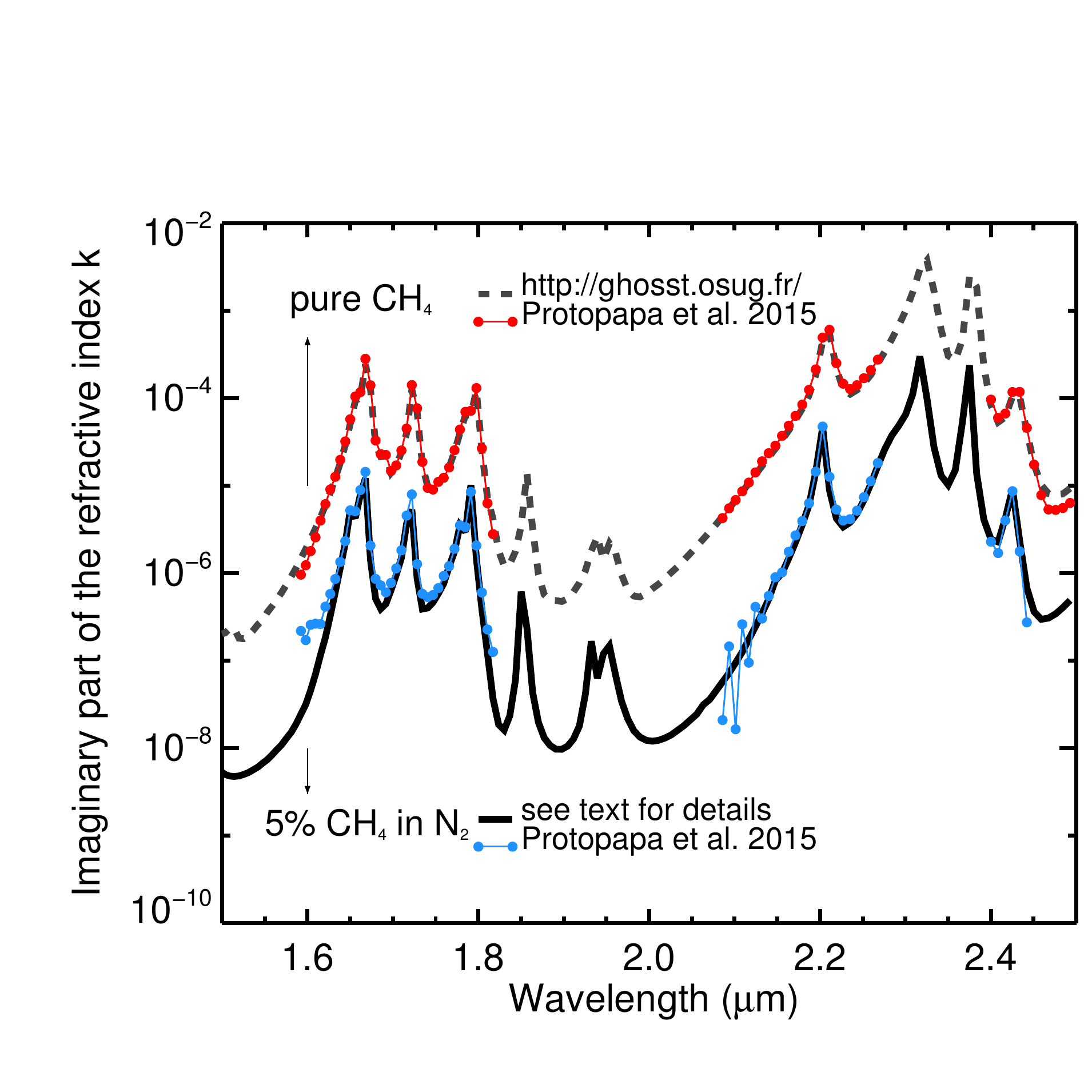}
	\fi
	\caption{The comparison between the imaginary part of the refractive index, $k$, of a \phaseNCH~mixture with 5\% \methane~in \nitrogen~computed numerically (black line, see text for details) and measured in the laboratory by \citet[][blue dots]{Protopapa2015} is shown. For reference, two sets of k for pure \methane~at similar temperatures (gray dashed line and red dots) are shown.}
	\label{figure1}
\end{figure}
\subsection{\nitrogen-rich and \methane-rich saturated solid solutions}
\citet{Protopapa2015} present optical constants for solid solutions of methane diluted in nitrogen (\nitrogen:\methane) and nitrogen diluted in methane (\methane:\nitrogen) at temperatures between 36 and 90~K, at different mixing ratios \url{http://www2.lowell.edu/users/grundy/abstracts/2015.CH4+N2.html}. This set of measurements, which includes optical constants of \nitrogen-rich (\phaseNCH) and \methane-rich (\phaseCHN) saturated solid solutions, cover the same wavelength range as the LEISA spectrometer. However, they are available in narrow blocks of wavelengths covering regions of intermediate absorption only. This implies that no optical constants are available for \phaseNCH~and \phaseCHN~around 1.5 and 2.0 \micron, where water ice, which turned out to be an undisputed component on the surface of Pluto, presents diagnostic absorption features.  While further efforts are being conducted to complete this data set, we adopt optical constants of pure \methane~as a proxy for \phaseCHN~and generate optical constants for \phaseNCH~following the method described by \citet{Doute1999}. 

The solubility limit of \nitrogen~in \methane~is 3\% at 40K. The wavelength shift of the \methane~bands in such a system is smaller than the LEISA resolving power -- shifts on the order of $\sim$2$\times$10$^{-4}$~\micron~are reported by \citet{Protopapa2015}, which is over an order of magnitude smaller than the LEISA resolution. Therefore, we do not expect our analysis to be significantly affected by the use of pure \methane~in place of \phaseCHN. This is still valid at temperatures lower than 40~K, as the solubility limit of  \nitrogen~in \methane~decreases with decreasing temperatures \citep[see the \methane-\nitrogen~binary phase diagram by][]{Prokhvatilov1983} and smaller shifts correspond to higher \methane~abundances. We use the optical constants of crystalline \methane-I at 39~K from \url{http://ghosst.osug.fr/} (see Table~\ref{Table2} for details). 
We follow the method described by \citet{Doute1999} to numerically generate optical constants for \phaseNCH. We consider two reference components: the first one is identified with pure \nitrogen~ice, and the second is \methane~initially diluted in the nitrogen matrix, but artificially normalized to unit concentration \citep{Quirico1999}. \citet{Doute1999} assume that, for each wavelength, the scalar product of the complex indices of \nitrogen~and the diluted \methane~by the vector \Bigg[1-$F^{\mathrm{CH_{4}}}_{\mathrm{\bf{\overline{N_{2}}}:CH_{4}}}$, $F^{\mathrm{CH_{4}}}_{\mathrm{\bf{\overline{N_{2}}}:CH_{4}}}$\Bigg] gives the optical constants of a mixture with a percentage of \methane~in solid \nitrogen~equal to $F^{\mathrm{CH_{4}}}_{\mathrm{\bf{\overline{N_{2}}}:CH_{4}}}$. We validate this numerical approach against the optical constants measured by \citet{Protopapa2015} for $F^{\mathrm{CH_{4}}}_{\mathrm{\bf{\overline{N_{2}}}:CH_{4}}}$=5\%, which is the solubility limit of \methane~in \nitrogen~at 40~K. We use the optical constants for pure \nitrogen~and \nitrogen:\methane~listed in Table~\ref{Table2}. As shown in Figure~\ref{figure1}, the numerical approach (black line) well reproduces not only the strengths of the measured (blue dots) \methane~bands but also that of \nitrogen~at 2.15~\micron. The concentration of \methane~in \nitrogen~($F^{\mathrm{CH_{4}}}_{\mathrm{\bf{\overline{N_{2}}}:CH_{4}}}$) is a free parameter in our study. 
\subsection{\water-ice}
Laboratory measurements show that infrared water ice absorption bands at 1.5, 1.65 and 2.0~\micron~change position and shape as a function of phase (crystalline or amorphous) and temperature \citep{Grundy1998}. We do not solve in this analysis for \water-ice temperature and phase and use instead optical constants of crystalline hexagonal water ice at 40~K (see Table \ref{Table2}, \url{http://www2.lowell.edu/users/grundy/abstracts/1998.H2Oice.html}). 
\begin{figure*}[!th]
	\ifincludegraphics
 	\centering
	\includegraphics[width=0.7\textwidth]{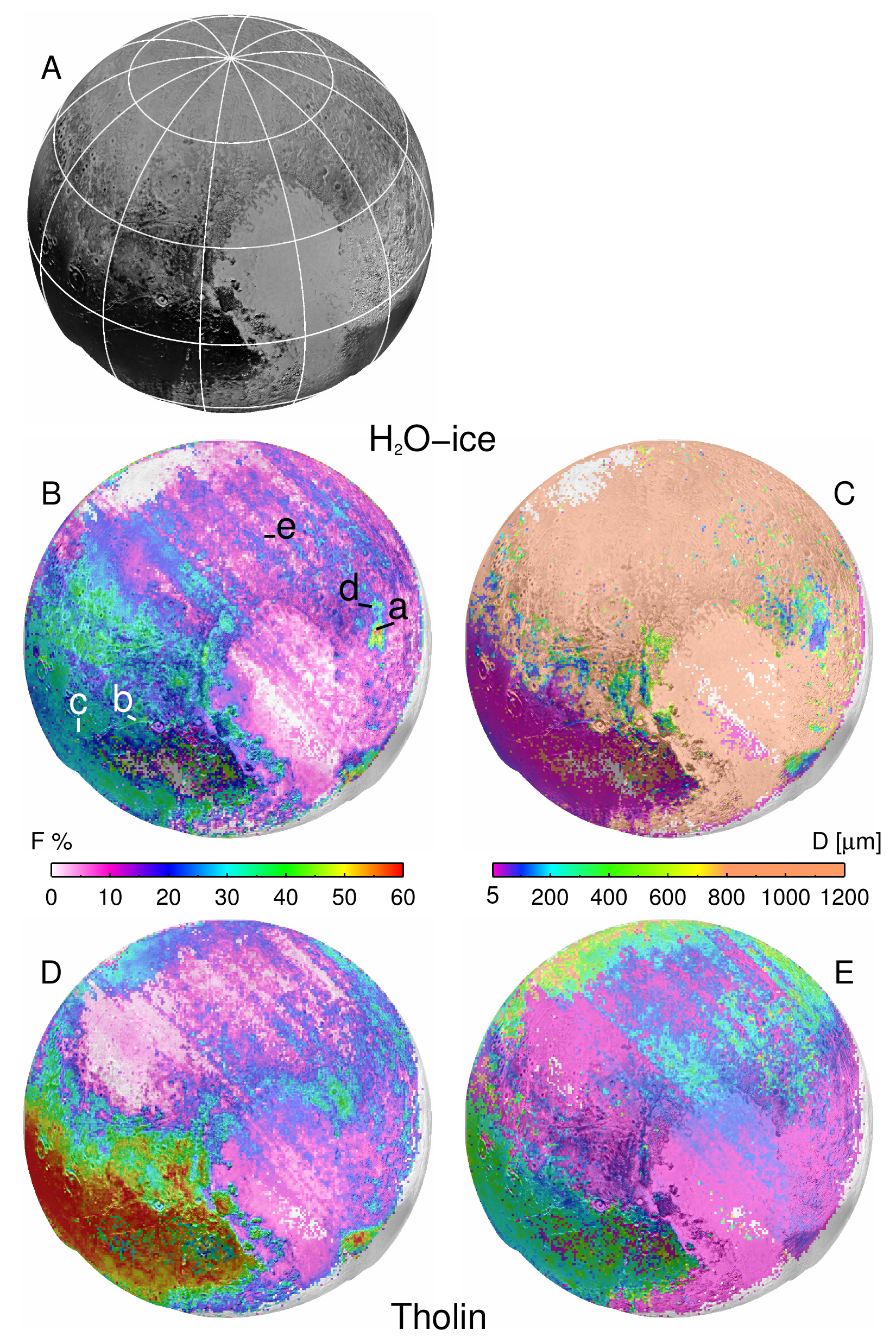}
	\fi
	\caption{Panel A shows the LORRI base map reprojected to the geometry of the LEISA observation.	
	Panels B and C show the pixel-by-pixel modeling results for \water-ice abundance, F\%, and effective diameter, D[$\mu$m], superposed on the LORRI base map. Areas of interest discussed in the text are labelled in panel B. Panels D and E show similar maps for Titan tholin. Regions with abundance lower than 10\% (white and purple) are below the noise level and therefore should not be overinterpreted.}
	\label{Figure2}
\end{figure*}
\section{Model discussion}\label{Model Discussion}
Applying the pixel-by-pixel modeling analysis described in Section~\ref{Spectral Modeling} is highly computationally expensive. To overcome such limitation, we have implemented our code with multi-threading capabilities. The main code splits the modeling of different surface units across Pluto between multiple threads of execution. We run our simulations on a multi-processor machine where these threads can execute concurrently. Furthermore, to speed up the task,  the pixel-by-pixel modeling analysis is
applied to the LEISA data degraded to a spatial resolution
of $\sim$12 km/pixel (for a total of $\sim$4$\times$10$^{4}$ pixels to model). Using 15 threads, the modeling of each LEISA scan requires approximately 10 hours.

We present results obtained by applying the Hapke radiative transfer model, assuming an areal mixture of the single endmembers. Several scattering theories \citep[\textit{e.g.},][]{Shkuratov1999,DouteSchmitt1998} as well as different types of mixtures (\textit{e.g.,} intimate) exist. These alternative methods could possibly provide similar quality of fits to the data as those presented here but with different percentages and grain size of the components \citep{Poulet2002}. However, we decided to interpret the Pluto New Horizons data with a simple model since it provides a reasonable fit to all 10$^{4}$ Pluto spectra (see Section \ref{Results}). The one presented here is an automated process and a perfect match between
observations and the model is beyond the scope of
the analysis. Ultimately, the goal is to perform a comparative study between Pluto's main surface units, which can only be done when the same modeling strategy is applied successfully to the full surface of Pluto.

As noted by \citet{Barucci2008}, grain size and abundance are sometimes entangled. This occurs mainly when the absolute I/F is unknown and when the analysis is not conducted over a wide wavelength range,  which fortunately is not the case for the New Horizons data. We acknowledge that this problem could still arise in absence of defined absorption bands, as in some areas dominated by non-volatile components.

The estimates of the concentration and particle size of each surface compound rely on the choice of the Hapke parameters $\xi$, $h$, $B_{0}$, and  $\overline\theta$, since these surface photometric properties affect the entire reflectance spectrum. As discussed in Section \ref{Spectral Modeling}, we treat these properties  as global quantities, constant across all of Pluto's terrains.  Given the high degree of surface variations on Pluto, this approximation may not be correct. Work is ongoing to determine the values of these parameters across Pluto's main surface units by inverting the New Horizons Multi-spectral Visible Imaging Camera \citep[MVIC;][]{Reuter2008} data acquired at different phase angles  \citep{Buie2016}. Our analysis will be revisited when the problem of Pluto's photometric properties will be unraveled, which may take some time given its complexity.

While New Horizons confirmed the detection of CO ice on Pluto's surface, mainly in Sputnik Planitia \citep{Grundy2016}, we did not include this component in our models. The main reason is the lack of optical constants in the continuum region, outside of the CO ice absorption bands. Assumptions on the absorption coefficient of CO ice in the continuum region would affect the determination of the abundance and grain size of the other surface compounds, for which optical constants are instead known. We will investigate the inclusion of CO ice in our models as well as other minor species like C$_{2}$H$_{6}$ in future work.
\section{Results}\label{Results}
We discuss below the spatial distribution of the abundance and grain size of non-volatiles and volatiles.  
\subsection{Non-Volatiles}
\begin{figure}[!th]
	\ifincludegraphics
 	\centering
	\includegraphics[width=0.43\textwidth]{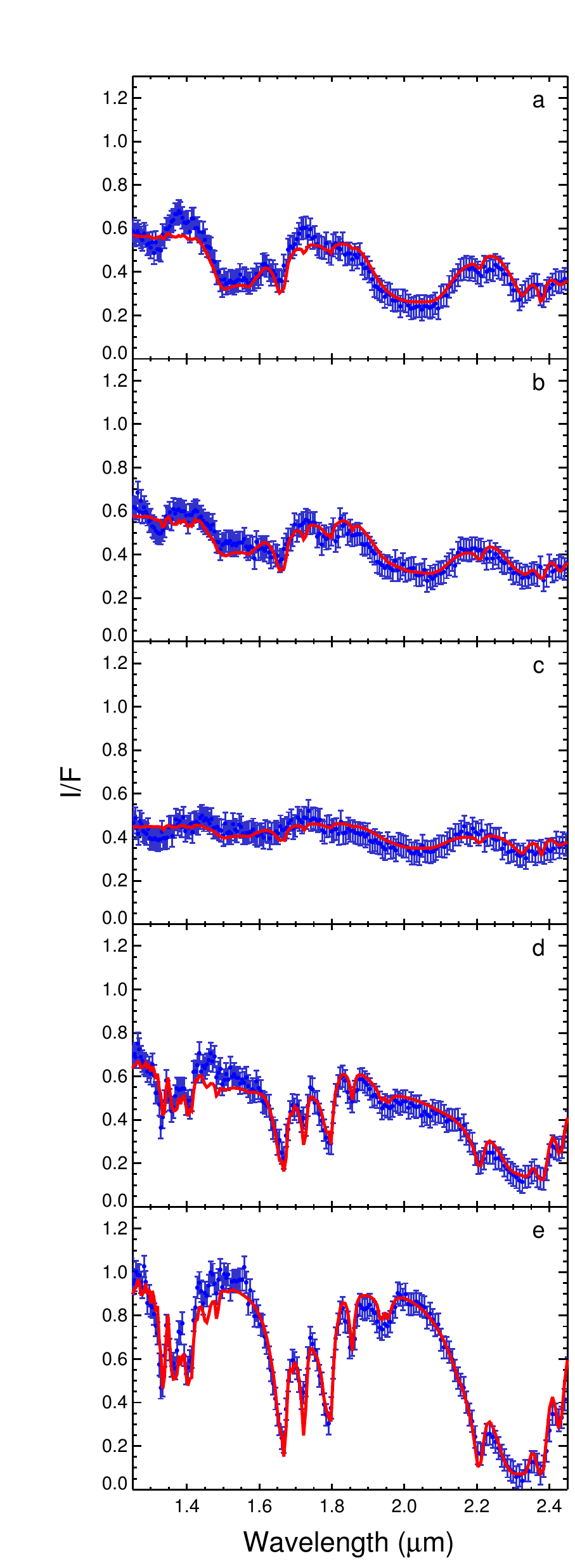}
	\fi
	\caption{Spectra extracted in regions labelled in Figure~\ref{Figure2} are shown as blue dots with 1$\sigma$ error estimates, and compared with their corresponding best fit models (solid red line).}
	\label{Figure3}
\end{figure}
Locations enriched in Titan tholin and \water~ice are correlated with specific geologic regions \citep[see panels B and D of Figure~\ref{Figure2}, \eg, Cthulhu Regio, Bar\'{e} Montes, Piri Planitia, Inanna Fossa,][]{Moore2016}. 
A change in \water-ice abundance and grain size (panels B and C of Figure~\ref{Figure2}) is observed across these different terrains. In Figure~\ref{Figure2} areas of interest are labelled and their corresponding spectra (blue points, with 1$\sigma$ error estimates), compared with the best fit modeling (red line), are displayed in Figure~\ref{Figure3}. The spectra are arranged in decreasing \water-ice abundance from top to bottom.

The region in the vicinity of Pulfrich crater (region ``a") stands out for being particularly enriched in \water~ice, with abundances close to 60\% (see Table \ref{Table3}). The enrichment of \water~ice in the vicinity of Pulfrich crater is in
good agreement with the map displaying the correlation coefficient between each LEISA spectrum and a template Charon-like \water~ice spectrum \citep{Grundy2016,Cook2016}. The spectrum extracted in this region (panel ``a", Figure~\ref{Figure3}) shows indeed strong diagnostic absorption bands due to \water~ice at 1.5 and 2.0~\micron, including the 1.65-\micron~feature, which is characteristic of the crystalline phase of the \water~ice \citep{Grundy1998}. The relative strength and shape of the \water-ice absorption bands are consistent with grain sizes on the order of $\sim$150~\micron. Notice that the best fit modeling suggests the presence not only of \water~ice but also of tholin and  \methane. The latter produces the absorption bands beyond 2.2~\micron. Differences between observations and modeling around 1.4 and 1.7~\micron~are attributed to flat-fielding, which still requires further optimization. We conclude that across Pluto's encounter hemisphere, no regions of 100\% \water~ice are observed, at this spatial resolution. 

The spectrum extracted along Virgil Fossae (region ``b") displays a spectral behavior, including the absolute value of $I/F$, similar to that of Pulfrich crater but with shallower \water-ice absorptions. The comparison of the modeling parameters obtained for these two terrains highlights the importance of radiative transfer modeling of the absorption bands to disentangle the effects of grain size, abundance and illumination geometry (see Table \ref{Table3}). The qualitative comparison between these two spectra (``a" and ``b") would lead to the conclusion that the fractional abundance of \water~ice is higher in Pulfrich crater than in Virgil Fossa. While this is still the case, a difference in the particle diameter of \water~ice is also observed. This is required to take into account the variation in illumination geometry, which plays as important a role as composition and texture in determining the observed spectral properties of the surface. 

The spectrum extracted in the Cthulhu Regio (region ``c") presents, with respect to the two spectra discussed above, shallower \water-ice absorption bands and a neutral spectral slope. This is consistent with smaller \water-ice particle diameters ($\sim$10~\micron). Furthermore, the absolute value of $I/F$ is lower (on the order of 0.4 instead of 0.6), consistent with a higher abundance of tholins. The spectral effect of tholins is indeed to darken the albedo level, without introducing any significant spectral features. Notice that the absolute $I/F$ values of Titan tholin decrease with increasing grain size. This explains our modeling results, which show regions across Pluto with lower reflectance values (\eg, Cthulhu Regio) corresponding to higher abundances of tholin and larger particle diameters (panels D and E in Figure~\ref{Figure2}). 

A few areas across Hayabusa Terra (\eg, region ``d") exhibit evidence of \water~ice. The strong \methane~absorption features not only beyond 2.2~\micron~but also across the full wavelength range are indicative of a greater contribution of this ice compared to the previously described regions. At the same time, the presence of \water~ice is highlighted by the depression in the spectral region around 2.0 \micron. These features would not be evident in a spectrum mainly dominated by methane ice, as in the case displayed in panel ``e'' of Figure~\ref{Figure3}. It is important to point out that spectra from near the north pole and from most of Tombaugh Regio do not show any spectral evidence for the presence of \water~ice. On the other hand,  the modeling results indicate abundances on the order of 10\% and corresponding grain sizes larger than 1000 \micron. Synthetic spectra of \water~ice with such large particles present saturated bands and very low reflectance values. Therefore we infer that water ice in this case is a filling material, spuriously adopted by the modeling code. This is supported by the fact that the model of region ``e'' with null \water~ice abundance, provides an equally good fit (comparable $\chi^{2}$) to the one presented in Figure~\ref{Figure2}. The absence of \water~ice mainly affects the grain size and abundance of the tholins, which in this instance have larger values than those reported in Table \ref{Table3} for case ``e''. Analogously, the contribution of \phaseNCH~in all cases where its abundance is lower than 10\% is considered not significant.
\begin{table*}[!bh]
\caption{Models}
\resizebox {\textwidth}{!}{%
\begin{tabular}{c | cc| cc | cc | cc | cc | c | c}
\hline
Region & \multicolumn{2}{c}{geometry}&\multicolumn{2}{c}{\water} & \multicolumn{2}{c}{Titan tholin} & \multicolumn{2}{c}{\phaseCHN} & \multicolumn{2}{c}{\phaseNCH} & $F^{\mathrm{CH_{4}}}_{\mathrm{\bf{\overline{N_{2}}}:CH_{4}}}$ & $\chi^{2}$\\
\hline
&$i[deg]$&$e[deg]$& F[\%]&D[\micron] &F[\%]&D[\micron] &F[\%]&D[\micron] &F[\%]&D[\micron]& [\%]&\\
\hline
a&64&49&60&147&19&37&21&82&0&n/a&n/a&0.78\\
b&39&35&37&71&48&142&15&376&0&n/a&n/a&1.44\\
c&55&58&31&11&58&268&11&73&0&n/a&n/a&0.76\\
d&62&48&26&172&9&38&57&794&8&181252&3.04&1.07\\
e&34&38&9&3058&12&68&60&695&19&55998&1.61&1.76\\
f&41&56&8&1270&12&165&53&682&27&62002&5.00&1.86\\
g&45&67&14&1708&26&424&15&247&45&131197&0.41&2.61\\
h&36&46&21&46&32&446&43&1137&4&10$^{7}$&0.10&2.34\\
i&57&35&10&2344&9&12&37&1022&43&590952&0.35&1.35\\
j&31&10&6&1907&18&52&52&941&23&200581&0.53&2.01\\
\hline
\end{tabular}}
\label{Table3}
\end{table*}
\subsection{Volatiles}\label{Volatiles}
\begin{figure*}
	\ifincludegraphics
 	\centering
	\includegraphics[width=\textwidth]{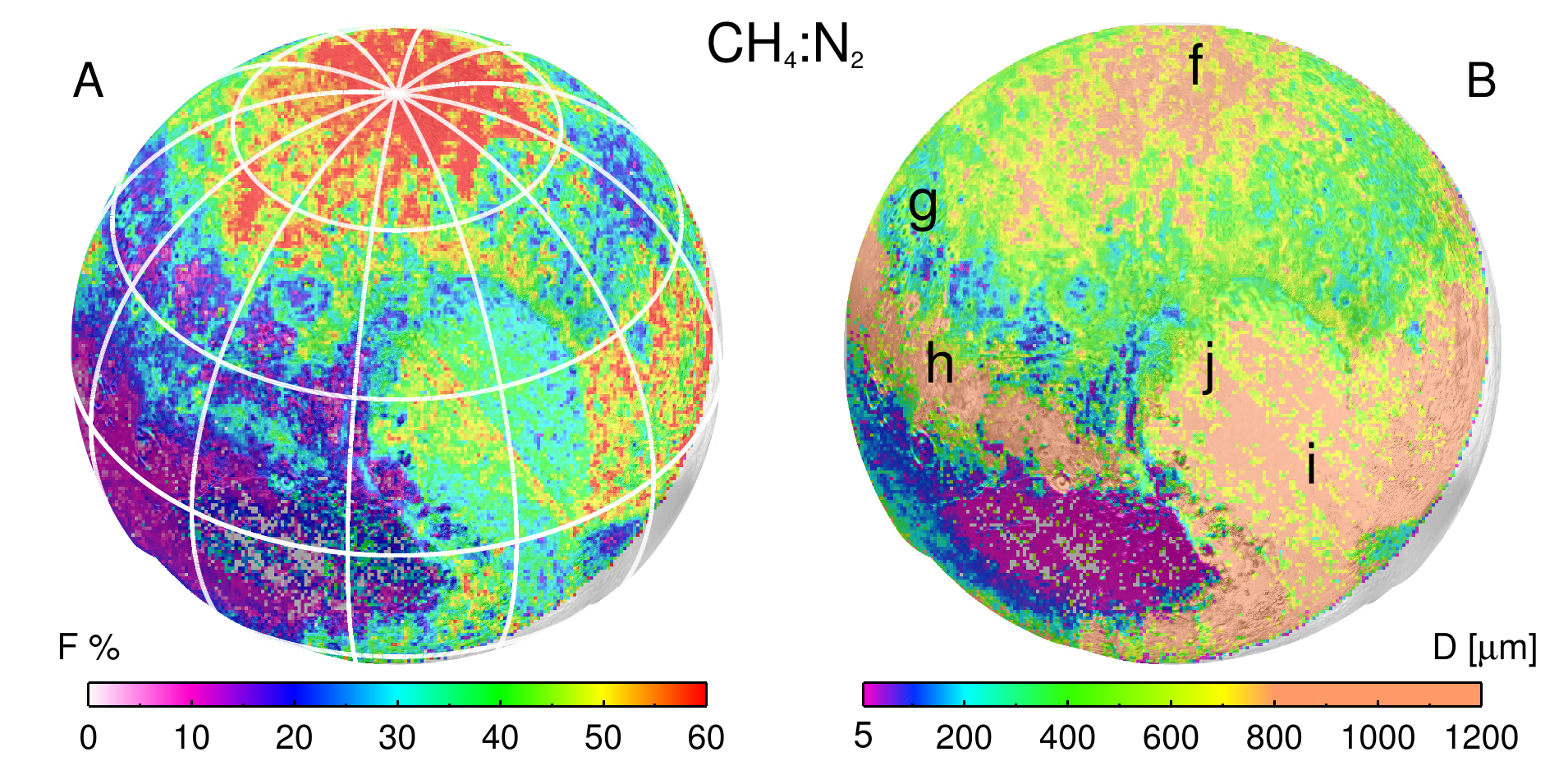}
	\fi
	\caption{Panels A and B show abundance and effective diameter of the \methane-enriched component, respectively,  as obtained from the pixel-by-pixel modeling analysis. The composition maps are superposed on the reprojected
LORRI base map. Areas of interest discussed in the text are labelled in panel B.}
	\label{Figure4}
\end{figure*}

\begin{figure*}
	\ifincludegraphics
 	\centering
	\includegraphics[width=\textwidth]{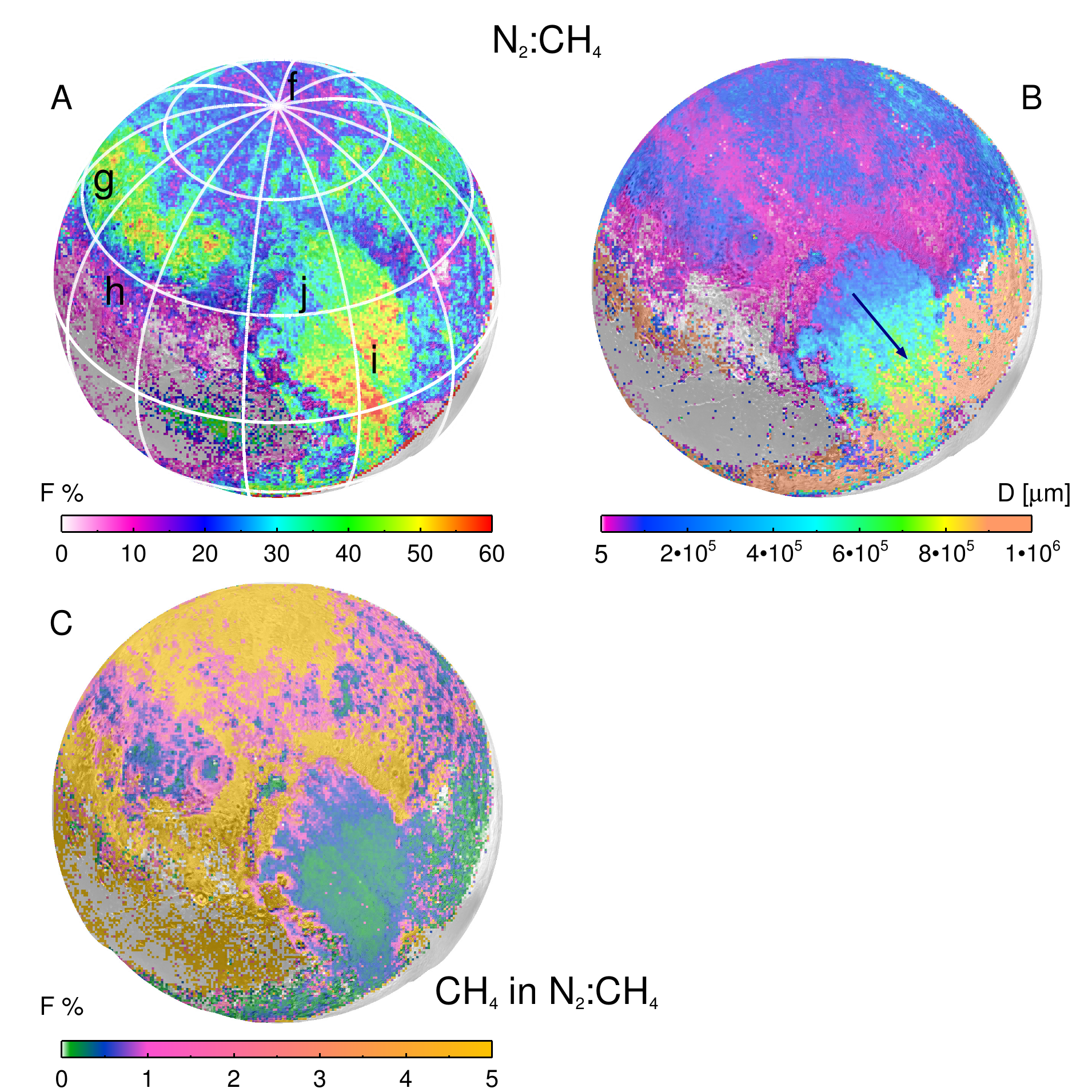}
	\fi
	\caption{Modeling results relative to abundance and path length of the \nitrogen-enriched component are shown in panels A and B, respectively. Panel C shows the dilution content of \methane~in \nitrogen. The composition maps are superposed on the reprojected
LORRI base map. Areas of interest discussed in the text are labelled in panel A. The arrow in panel B indicates the direction of the \nitrogen~sublimation transport discussed in the text.}
	\label{Figure5}
\end{figure*}

No major region across Pluto's surface is completely depleted of \methane~(panel A, Figure \ref{Figure4}). \nitrogen~on the other hand presents a more localized distribution (panel A, Figure \ref{Figure5}). We identify in the volatile maps large scale latitudinal regions, which differ in abundance and texture of the \methane-rich and \nitrogen-rich components. 

\begin{figure}
	\ifincludegraphics
 	\centering
	\includegraphics[width=0.45\textwidth]{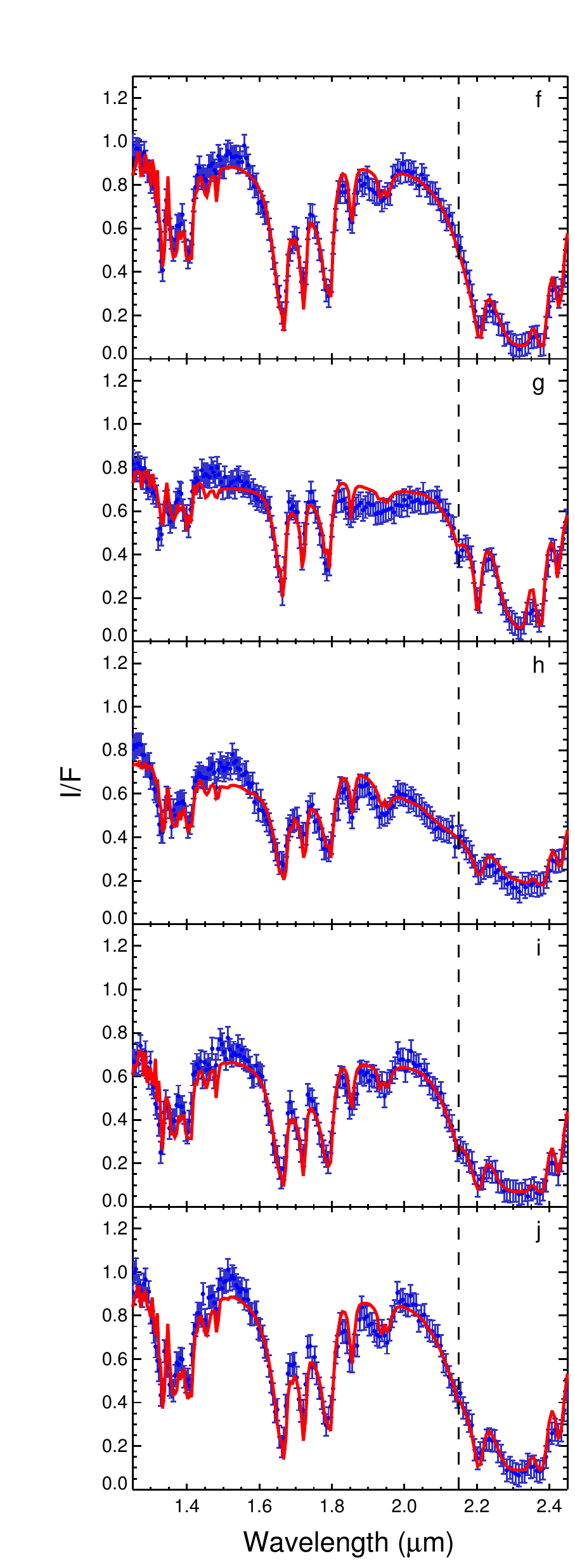}
	\fi
	\caption{Spectra extracted in regions labelled in Figures~\ref{Figure4} and \ref{Figure5} are shown as blue dots and compared with their corresponding best fit modeling (solid red line).}
	\label{Figure6}
\end{figure}

The first of these terrains is Lowell Regio, labelled ``f" (Figure~\ref{Figure4}-\ref{Figure5}). The second is the belt stretching from latitude 35$^{\circ}$N to 55$^{\circ}$N and crossing Burney Crater and Hayabusa Terra (identified with ``g" in Figure~\ref{Figure4} and \ref{Figure5}). This belt appears to be interrupted by Sputnik Planitia, a deep reservoir of convecting ices \citep{McKinnon2016,McKinnon2016Natur,Trowbridge2016}. The relative amount of the \methane-to-\nitrogen-rich components decreases with decreasing latitude. This is demonstrated by the spectra extracted from representative locations in these two surface units (Figure \ref{Figure6}). Spectrum ``f", contrary to its counterpart  ``g", does not display a \nitrogen-absorption feature at $\sim$2.15~\micron~(vertical dashed line). The modeling indicates a \nitrogen~distribution on the order of 30\% and 45\% across Lowell Regio and between 35$^{\circ}$N and 55$^{\circ}$N, respectively (see Table~\ref{Table3}). Notice that the presence of an \nitrogen-rich component is marked not only by the $\sim$2.15~\micron~absorption band but also by the absolute level of the continuum, being higher where \nitrogen~occurs. Furthermore a difference in the path length of the \nitrogen-rich component is observed, being larger at lower latitudes. This is justified by the \nitrogen~feature getting deeper with larger grains. 

Using the MVIC \methane~equivalent-width map, \citet{Grundy2016} identified another region of interest at low-latitudes at the border of Cthulhu Regio and to the east of Tombaugh Regio, embracing Tartarus Dorsae. This area was identified by a strong 0.89-\micron~absorption band, possibly due to a higher \methane~abundance and/or to
especially large particle sizes. Our study points out the same area of interest (panel B, Figure~\ref{Figure4}, region ``h") and attributes its peculiarity to particularly large particle sizes  ($\sim$1000~\micron, see Table~\ref{Table3}). While in MVIC observations the diagnostic is the equivalent width of the 0.89-\micron~\methane~band, LEISA observations stand out for the broadness of the 2.3-\micron~\methane~band (panel ``h'', Figure \ref{Figure6}). 

Existing literature on modeling of ground-based near-infrared spectroscopic measurements of Pluto highlights the presence of two ``pure" \methane~components with different grain sizes. The first one is characterized by large mm~grains, while the other is described by smaller grains on the order of $\sim$100~\micron. This result is obtained when modeling Pluto spectra with intimate \citep{Merlin2015} as well as areal mixtures \citep{Protopapa2008}.  However, when modeling LEISA spectral data, this distribution of grain sizes is not necessary at the pixel level. We attribute this behavior to the high spatial resolution of the \NH~data. In fact, the grain size map of the \methane-rich component presented in this paper shows the presence of regions with particles spanning a range from $\sim$70~\micron, to $\sim$1000~\micron, justifying the models of the ground-based measurements of Pluto. 

The latitudinal pattern described so far is interrupted by Sputnik Planitia, a region set apart by strong 2.15-\micron~absorption diagnostic of substantial \nitrogen~contribution (region ``i" in Figure~\ref{Figure5}). This conclusion was reached already by \citet{Grundy2016} from the analysis of the equivalent width of the 2.15-\micron~band. However, this is only a rough approximation for the
\nitrogen~ice abundance. In fact this absorption band is a minor feature on the wings of the 2.2-\micron~\methane~band and it is only through modeling that its contribution can be isolated and quantified. Three free parameters in our modeling all point to the greater role that \nitrogen~plays in this part of Pluto than anywhere else on the encounter hemisphere. They are the large abundance of \phaseNCH~(panel A, Figure~\ref{Figure5}), the small amount of \methane~diluted in \nitrogen~(panel C, Figure~\ref{Figure5}), and the large path length (or equivalently particle diameter $D$, panel B, Figure~\ref{Figure5}). Compositional differences are observable within Sputnik Planitia (regions ``i'' and ``j'' in Figure~\ref{Figure5}): the northwest part of Sputnik Planitia presents, with respect to the center, a shallower \nitrogen~absorption band, and therefore a smaller abundance of \phaseNCH~with higher concentrations of \methane~in \nitrogen, and smaller path length (Table~\ref{Table3}). This dichotomy is suggestive of a possible transport of \nitrogen~from the northwest to the
south of Sputnik Planitia (see arrow in Figure~\ref{Figure5}, panel B). It is likely due to sublimation and redeposition driven by the north-south insolation gradient. Sublimation of \nitrogen~in the northwest part of Sputnik Planitia leaves behind \methane, which is less volatile. This would justify the larger amount of \methane~in the \nitrogen-rich component, as well as the shrinkage of grains. The larger abundance of \phaseNCH~at the center of the basin is instead consistent with condensation of \nitrogen.

\section{Discussion}
We present a quantitative analysis of the spatially resolved spectral maps of Pluto's surface obtained with the \NH~Ralph/LEISA instrument. This analysis provides constraints on the amount and grain size of Pluto's surface components. The resulting spatial distribution of ices and tholin appears to be mostly consistent with the qualitative analysis performed on the same data set by \citet{Schmitt2016} using different techniques. 

Simplifying the compositional findings into rough latitudinal regions, the maps that
we have presented in Figs. \ref{Figure2}, \ref{Figure4} and \ref{Figure5} give the following broad
picture. Straddling Pluto's equator, Cthulhu Regio is a band of low
reflectance tholin-like material with very little \methane~or \nitrogen~
signatures. To the north at about 20$^{\circ}$ latitude this
transitions to a region of \methane~rich material, which in turns yields
to an \nitrogen~and \methane~mixture at about 35$^{\circ}$ latitude. Further
to the north, by about 55$^{\circ}$ latitude, the \nitrogen~signature
smoothly tapers off to an expansive polar plain of predominantly
\methane~ices.

Pluto's intriguing diversity of surface features traces back to its
size, which given cold temperatures in the outer Solar System, is
sufficient to retain a thin atmosphere. Seasonal deposits and more
substantial reservoirs of \methane~and \nitrogen~ices cover the surface
which, in the absence of volatiles, would be expected to be composed
principally of water ice \citep{Schaller2007}. Our compositional bands are composed of volatile materials and roughly follow lines of latitude, which suggest that insolation is likely controlling the distribution. Conversely, the fact that the boundaries are somewhat ragged and not perfectly aligned with latitude indicates that other effects, likely including surface composition and topography, are important too. We neglect these local effects here, instead concentrating on the global structure apparent in our maps.

\begin{figure}
	\ifincludegraphics
 	\centering
	\includegraphics[width=0.7\textwidth]{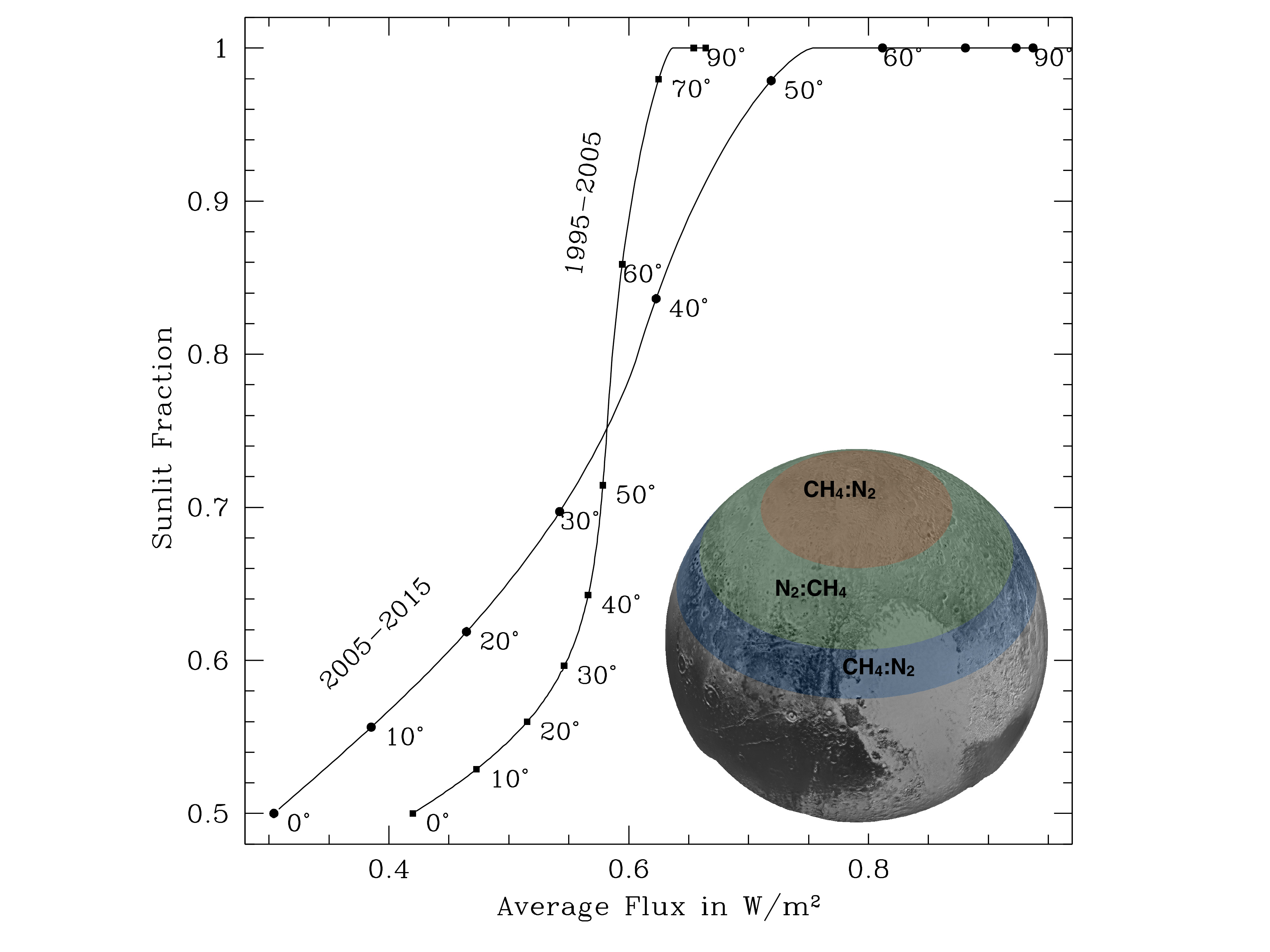}
	\fi
	\caption{The fraction of a Pluto rotation period that a point at a given latitude is sunlit versus average flux in W/m$^{2}$. Northern latitudes are marked with symbols spaced at 10$^\circ$ intervals. The filled squares show the average energy flux to Pluto's northern hemisphere over the decade 1995-2005 while the filled circles show the same for the past ten years. Pluto's north pole has been sunlit since 1987 and, over the past 20 years, has received more solar illumination than any other latitude. Differences are relatively minor around the turn of the century, but become much more pronounced after 2005. The intense heating correlates well with polar depletion of \nitrogen~(see the reddish circle on the Pluto inset).}
	\label{Figure7}
\end{figure}

The transitions that we find with our maps of volatile cover correlate
well with expectations of vigorous spring sublimation after a long
polar winter. The Sun returned to illuminate Pluto's north pole in
late 1987 and for the past 20 years has been depositing more energy at
polar latitudes than in the temperate zone. Continuous
illumination northward of 75$^{\circ}$ over the past twenty years, and
northward of 55$^{\circ}$ over the past ten years (Figure~\ref{Figure7}, sunlit fraction equals one), seems to have
sublimated the most volatile N$_2$ into the atmosphere, with the best
chance for redeposition occurring at points southward. This loss of
surface \nitrogen~appears to have created the polar bald spot seen most
clearly in Figure \ref{Figure5} (panel A), shown schematically in Figure \ref{Figure7}, and also predicted by \citet{hansen96}. This sublimation front is advancing southward ever more rapidly as illumination
in the northern hemisphere strengthens. 

Regions southward of
35$^{\circ}$ have been exposed to strong solar
illumination with average fluxes greater than 0.4 W/m$^2$ for $\sim$40 years, twice as long as the polar regions (compare with Figure~\ref{Figure7}). From 1975-1995, these low latitude regions (blue band in Figure \ref{Figure7}) experienced much stronger heating than regions to the north, leading to sublimation of \nitrogen~and redeposition of \nitrogen~ice to points northward. We argue that a slowly moving \nitrogen~sublimation front has likely expanded northward from Cthulhu Regio and is destined to meet the more rapidly southward moving polar sublimation front within the decade.

The \nitrogen-rich component
dominates mainly in Sputnik Planitia. The ices in this reservoir are the result of long term processes \citep{Spencer1997, Earle2015DPS, Earle2016, Hamilton2016, Bertrand2016}, unlike the seasonal latitudinal bands discussed above. For a plausible Pluto surface temperature
of 40~K, the \nitrogen-\methane~binary phase diagram
shows that, in the case of thermodynamic equilibrium and if both the \methane-rich and \nitrogen-rich components are present, the solubility limit of \methane~in \nitrogen~is about 5\%. However,
setting this as a constraint failed to reproduce the
strong 2.15-\micron~\nitrogen~band. If we let the concentration of \methane~in \nitrogen~vary, we find that
amounts close to 0.3\% are needed to reproduce the
2.15-\micron~\nitrogen~band. It is not the first time that such low concentrations of \methane~in \nitrogen~are adopted in modeling analysis to reproduce Pluto's spectrum (\eg, \citet{Doute1999} obtained 0.36\% of \methane~diluted in \nitrogen~in their best fit model considering geographical mixture). Because we have a finite \methane-rich component, this finding could be interpreted as a sign of a much lower temperature in Sputnik
Planitia, in agreement with that reported by \citet{Hamilton2015}. However, a value of 0.36\% of \methane~diluted in \nitrogen~would imply \nitrogen~to be in $\alpha$-phase, contrary to the evidence collected so far. Because of the uncertainties in the binary phase diagram and because the contribution of CO ice was not taken into account, we leave the temperature interpretation  to a further effort which will include the analysis of the 2.1-2.25~\micron~LEISA segment. Furthermore we intend to explore different modeling approaches, including layering. In fact, the temperature interpretation stands only if the \methane-rich component is locally present with the \nitrogen-rich one. However, a dynamical differentiation between \nitrogen-rich and \methane-rich components could occur during sublimation producing possible layering \citep{Stansberry1996}. The physical properties of \nitrogen~across Sputnik Planitia could in fact suggest a possible sublimation transport of this volatile from the northwest to the center.

\section{Acknowledgments}

This work was supported by the New Horizons project. S. Protopapa gratefully thanks the NASA Grant and Cooperative Agreement for funding that supported this work  (grant
\#NNX16AC83G). B. Schmitt, S. Philippe, and E. Quirico acknowledge CNES that supported this work. Simulations were performed on the YORP cluster administered by the Center for Theory and Computation, part of the Department of Astronomy at the University of Maryland. We thank the free and open source optical constants repositories  for empowering us with key tools used to complete this
project. We thank Dr. F.~E. DeMeo and an anonymous reviewer for comments that helped to improve the paper. Finally, S. Protopapa thanks Dr. M. S. P. Kelley and Dr. G. Villanueva for useful discussions.



\section*{References}
\bibliographystyle{icarus} 
\bibliography{Protopapa_Pluto_NewHorizons_Bibtex}





\end{document}